\documentstyle{laa}
\begin{document}
\thesaurus{03(11.03.1; 12.03.01; 12.12.1)}
\title{
The power spectrum implied by COBE and the matter correlation function}
\author{S.~Torres\inst{1}\thanks{On leave from
Universidad de los Andes, and Centro Internacional de
F\'{i}sica, A.A. 49490,
Bogot\'{a}, Colombia}
\and R.~Fabbri\inst{2}
\and R.~Ruffini\inst{1}}
\offprints{S.~Torres}
\institute{
ICRA (International Center for Relativistic Astrophysics),
Dipartimento di Fisica, Universit\`{a} di Roma 1,
Pzza. Aldo Moro 2, I-00185, Roma, Italy. e-mail: 40174::TORRES
\and Sezione di Fisica Superiore, Dipartimento di Fisica dell'Universit\`{a}
di Firenze, Via S. Marta 3, I-50139, Firenze, Italy}
\date{Received date; accepted date}
\maketitle

\begin{abstract}
A phenomenological power spectrum of primordial density perturbations has been
constructed by using both COBE data to probe the large wavelength region,
and a double power law,
locally deduced from galaxy catalogs,
which describes the matter correlation function up to
tens of Megaparsec. The shape of the spectrum $P(k)$ of density fluctuations
exhibits a peak that singles out a characteristic wavelength $\lambda_{\rm
peak}$ proportional to the cutoff radius $R_0$ in the matter distribution
(comparable to the distance at which matter becomes anticorrelated). From a
least squares fit to COBE's angular correlation function we obtain $R_0 =
35 \pm 12 h^{-1}$Mpc for the correlation length, and $n = 0.76 \pm 0.3$ for the
spectral index of $P(k)$ in the large wavelength region.
The inferred scale in
the spectrum is $\lambda_{\rm peak} = 51 \pm 18 h^{-1}$Mpc.
This number agrees with that derived from the analysis of the correlation
function of matter and with a preferred scale identified in IRAS PSC.
\keywords{Galaxies: clustering -- Cosmology: cosmic microwave
background -- large-scale structure of Universe}
\end{abstract}

\section{Introduction}

After the positive detection of anisotropies in the cosmic microwave background
radiation (CMB) by COBE-DMR (Bennett et al. 1992; Smoot et al. 1992),
it is now possible to directly probe the spectrum
of primordial density fluctuations (PDF). Early attempts
to place restrictions on the
spectrum of PDF relied mainly on the existing data on
large scale matter distribution or on theoretical prejudices, depending
on what region of the spectrum was explored.
While  matter distribution
is useful in determining the shape
of the `evolved' spectrum at relatively
large wave-numbers,
large angular scale CMB anisotropies
are sensitive to the primordial spectrum
at low wave-numbers.

Analysis of the QDOT IRAS surveys and the angular two-point
autocorrelation function $w(\theta)$ (Peacock 1991) shows that both
data sets
are consistent with a spectral function
$P(k)$ that approaches $k^{-1.4}$ for large $k$. However,
increasing the depth of the surveys has revealed the presence of a
break in $\xi(r)$ (the two-point
galaxy-galaxy correlation function; Calzetti et al. 1992),
the existence of a region where $\xi(r)$ is negative
(Guzzo et al. 1991), and also the possible
detection of a cutoff radius $R_0$ of order 30 $h^{-1}$ Mpc.
In this paper
we make use of the double power law in $1 + \xi(r)$  to derive
a properly normalized power spectrum for large $k$, and we
match it to a power law
$P(k) \propto k^n$ for small $k$. The matching point is selected so that
the power spectrum makes a smooth transition from the primordial
power law to the `evolved' spectrum. As a result, the peak in the
spectrum turns out to be of the same order of magnitude as $R_0$.
The cutoff radius
in the matter distribution $R_0$ and the spectral index $n$
are then the only two important quantities, and we can treat them as
free parameters. A calculation of anisotropies
in the CMB with this model for the power spectrum allows us to
fit these parameters to COBE's angular correlation function.
The quadrupole amplitude allows for an independent check for
consistency.

The fitting turns out to be very satisfactory and provides for $R_0$ a best
value of $35 \pm 12 h^{-1}$Mpc
very close to the optimal value inferred from observations
of the galaxy-galaxy correlation function (Calzetti et al. 1991).
The best value for the primordial spectral index is
found to be $0.76 \pm 0.3$,
with a central value somewhat lower than
predicted by standard
exponential inflation but, because our error bars, still consistent
with it and
with Smoot et al. (1992).
A relatively low index is not peculiar to the
model because for a pure power-law spectrum we find similar results.
With respect to this, we notice that
an attempt to fit COBE data to a power law spectrum of
density fluctuations within CDM models results in a spectral index
$0.6 < n < 1.0$ (Adams, et al. 1992). Kashlinsky (1992) also finds
$n<1.0$ in order to satisfy data on both the large scale galaxy distribution
and large scale streaming motions.

A feature which, on the other hand, is peculiar to our model is the close
relationship between the cutoff radius of the matter correlation
function and the peak in
the spectrum. As a matter of fact, the maximum of $P(k)$ is located at
$\lambda_{\rm peak} = 51 \pm 18 h^{-1}$Mpc. This number agrees
with the privileged
length scale of about $50 h^{-1}$Mpc found in the IRAS Point
Source Catalog by Fabbri and Natale (1993a). A further analysis (Fabbri
and Natale 1993b) provides evidence for the interpretation of this scale
as a peak in the product $k^{1/2} P(k)$. However, it does not agree with the
very large scales recently inferred from redshift surveys
(Fisher et al. 1993; Einasto et al. 1993;
Jing \& Valdarnini 1993). This
point will be discussed in Sect. 5.

\section{Power spectrum}

To construct a phenomenological spectrum we started from
the autocorrelation of the density fluctuation
$\delta \rho/\rho$, which is equal
(up to a delta function at the origin)
to the particle
correlation function $\xi(r)$ and can be
parametrized as follows (Calzetti et al. 1992):
\begin{equation}
1+\xi(x) = \left\{ \begin{array}{ll}
    A_1 x^{-\gamma_1}   & x < x_t \\
    A_2 x^{-\gamma_2}   & x_t < x \leq x_0 \\
    1                   & x > x_0
    \end{array}
\right. \end{equation}

To make the calculations independent of Hubble's constant
$H_0 = 100 h$~Km~s$^{-1}$~Mpc$^{-1}$ we adopt the
change of variable $x=hr$.
The correlation function admits four
independent parameters, which we choose to be
the spectral indexes $\gamma_1$ and $\gamma_2$ and the transition points
$x_t$ and
$x_0$. The amplitudes $A_1$ and $A_2$
are then calculated in terms of these parameters
by demanding continuity of $1+\xi(x)$ at $x = x_t$
and the normalization
condition on the correlation function
$\int_V \xi(r) r^{2} dr = 0$.
The resulting formulae for the amplitudes are:
\begin{equation}
A_2 = \frac {3-\gamma_2} {3} x_0^{\gamma_2}
\left[
1 + \frac{\gamma_1 - \gamma_2}{3-\gamma_1}
\left( \frac{x_t}{x_0} \right)^{3-\gamma_2}
\right]^{-1}
\end{equation}
\begin{equation}
A_1 = A_2 x_t^{(\gamma_1 - \gamma_2)}.
\end{equation}

Three parameters are determined
by the existing
data, namely $ \gamma_1 = 1.8, \ \gamma_2 = 0.8, \
x_t = 3.0$~Mpc. It turns out that our results are quite insensitive to
the precise values of $ \gamma_1 $ and $ x_t $, and moderate differences are
found varying $\gamma_2$ in the range $0.7-0.9.$ On the other hand, the
value of the cutoff radius $x_0 = hR_0$ is
important for our calculations because
the dependency of
$A_1$ and $A_2$ on this parameter  affects the
normalization of $P(k)$
and thereby the CMB anisotropies. We treat
$x_0$ as a free parameter since it is  still
poorly determined.

Let us now define the spectrum of PDF in terms of the rms density
fluctuations:
\begin{equation}
\left( \frac{\delta \rho}{\rho} \right)^2 =
\int P(k) \Phi_k^2(\eta) {\rm d}^3k,
\end{equation}
where $k$ is a
dimensionless wavenumber defined as:
\begin{equation}
k = \frac{4 \pi c}{H_0 \lambda} =
\frac{\underline{k}}{h \alpha},
\end{equation}
with $\underline{k}$ the physical comoving wavenumber,
$\alpha \equiv 1/6000$~Mpc$^{-1}$, and $\eta$ the conformal time.
The temporal
evolution of perturbations is described by
$\Phi_k(\eta)$, which in the linear regime
is independent of wavenumber, and
for $\Omega=1$ can be set equal to $\eta^2/10$ with $\eta_o = 1$.
The spectrum of PDFs, being the
Fourier transform of the matter correlation function
(Peebles 1980), results in
\begin{equation}
P(k) = \frac{\alpha^3}{2\pi^2\Phi^2(\eta_o)}
\int_0^{x_0} x^2 \xi(x) \frac{\sin(\alpha kx)}
{\alpha kx} {\rm d}x.
\end{equation}

The above integration was done numerically using a combination
of both the Clenshaw-Curtis and Gauss-Kronrod methods which are
adequate for an oscillating integrand (Piessens et al. 1983).
The results of the Fourier transform shows the presence of
a peak in $P(k)$ at a wavenumber $k_{\rm peak}$ inversely
proportional to
$x_0^{-1}$ ($k_{\rm peak} \approx 2.56 \times 10^4 \mbox{Mpc}/x_0$).
Beyond the peak the function
$P(k)$ shows an oscillatory behavior (due to the sharp edge cutoff of
$\xi(r)$)
at first, but eventually
reaches a power law behavior with a
negative slope $P(k) \propto k^{-1.2}$,
in agreement
with the results of Peacock (1991). For
small wavenumbers $k<k_{\rm peak}$
we find $P(k) \propto k^2$.
This asymptotic law would be found from any correlation function which is
forced to vanish beyond a maximum radius $x_0$, or decreases as $x^{-\gamma}$
with $\gamma > 5$, provided the integral of $\xi$ over space vanishes.
This behavior does not agree with COBE's data. However,
since the domain where
the galaxy-galaxy correlation function
appreciably differs from zero is limited to
relatively small scales,
$P(k)$ can be modified in the small-$k$ region without appreciably
affecting $\xi(x)$. Therefore we replaced the
$P(k) \propto k^2$ law by a power function with a spectral
index $n$
left as a free parameter to fit with COBE's data. In order to
minimize the perturbation to $\xi (x)$ we extrapolated the spectrum in Eq. (6)
down to the wavenumber $k_m$ where its slope assumes the desired value of $n$.
For $k < k_m$  the
power law function $P(k)=P_m (k/k_m)^n$ was adopted.
The parameter  $P_m$
is the value of the original, $\xi -$derived spectrum in $k = k_m$.
Continuity
of both the spectral function and its first derivative is therefore
guaranteed.
Figure 1
illustrates the form of the function $P(k)$ used here for
our best values of $x_0$, $n$ and for $\gamma_2 = 0.8 $.
We have found a power series solution to the Fourier transform in Eq. (6)
neglecting the region where $\xi(r)$ has slope $-1.8$:
\begin{eqnarray}
P(k) & = & \frac{200}{3 \pi^2} (\alpha x_0)^5 \gamma_2 k^2 \nonumber \\
     &   & \times \sum_{i=1}^{\infty} (-1)^{i+1}
\frac{i(i+1)}{2i+3-\gamma_2}
\frac{(\alpha x_0 k)^{2i-2}}{(2i+3)!},
\end{eqnarray}
which can be conveniently
used to find the matching point $k_m$
of the function an its first derivative. An approximation which is
sufficient to our purpose is
\begin{eqnarray}
k_m^2 & = & \frac{14}{(\alpha x_0)^2}
\frac{2-n}{4-n}
\frac{7-\gamma_2}{5-\gamma_2}   \nonumber \\
 & & \times \frac{1-\frac{7}{18}F(n)G(\gamma_2)}{1-\frac{7}{9}F(n)G(\gamma_2)}
\end{eqnarray}
\begin{equation}
F(n) = \frac{(2-n)(6-n)}{(4-n)^2}
\end{equation}
\begin{equation}
G(\gamma) = \frac{(7-\gamma)^2}{(5-\gamma)(9-\gamma)}.
\end{equation}

\begin{figure}
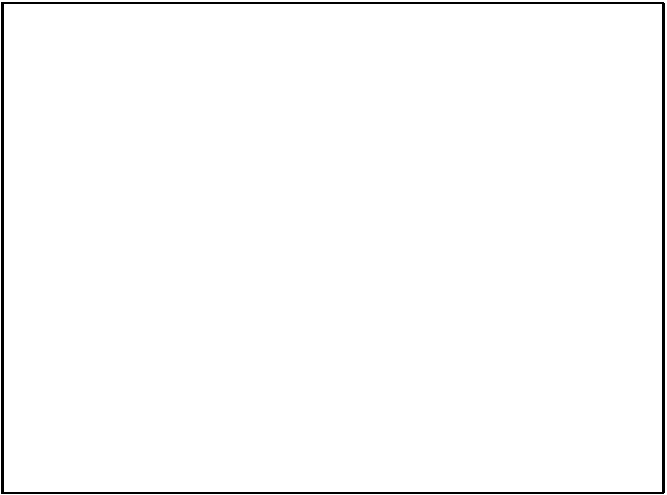

\picplace{6.5cm}
\caption[ ]{Spectrum of primordial density fluctuations for
$x_0 = 34.8$ Mpc, $n = 0.76$ and $\gamma_2 = 0.8$}
\end{figure}

Since Eqs. (6)-(8) do not contain $x_t$ and $\gamma_1$, they cannot describe
the large-$k$ region. However, this region is quite ineffective to influence
the CMB anisotropies.

\section{CMB Anisotropies}

With a properly normalized spectrum of PDF, one can
calculate the CMB large scale
fluctuations and angular correlation function.
The anisotropy field is
conveniently expressed in terms
of the coefficients $a_{\ell m}$ in a harmonic expansion:
\begin{equation}
\frac{\Delta T}{T_0} = \sum_{\ell \geq 2}^{\infty}
\sum_{m=-\ell}^{\ell}
a_{\ell m}Y_{\ell m}(\theta,\phi),
\end{equation}
with the dipole terms ($\ell = 1$)
excluded.
For random perturbation fields
the variance of $a_{\ell m}$
defines the angular power spectrum:
\begin{eqnarray}
\langle a_{\ell}^2 \rangle  & = & \sum_{m=-\ell}^{\ell}
\langle a_{\ell m}^2 \rangle \nonumber \\
             & = & (2\ell + 1)  \langle a_{\ell m}^2\rangle.
\end{eqnarray}

Experiments such as COBE-DMR which produce a sky map of the
cosmic background radiation give a direct measure of the
angular correlation function
$C(\theta , \sigma)$, which can be theoretically predicted by
\begin{equation}
C(\theta, \sigma) =
\frac{1}{4 \pi} \sum_{\ell}\langle a_{\ell}^2 \rangle
W_{\ell}^2 P_{\ell}(\cos \theta),
\end{equation}
where $P_{\ell}(\cos \theta)$ are the Legendre polynomials,
and the `filter' function for a Gaussian
beamwidth $\sigma = 1/(\ell_{0}+\frac{1}{2})$ radians is
\begin{equation}
W_{\ell} = \exp \left[-\frac{(\ell + \frac{1}{2})^2}
{2(\ell_0 + \frac{1}{2})^2} \right].
\end{equation}

For COBE-DMR filtering effectively starts at
$\ell_0 \approx 18$
(corresponding to an angular resolution of $3.1^\circ$), but in our case
terms up to $\ell \approx 60$ should
be included for an accurate calculation.

The imprint left on the CMB produced by metric perturbations
during the epoch of last scattering,
(Sachs \& Wolfe, 1967)
can be calculated by evaluating the integral:
\begin{equation}
a_{\ell}^2 =  \left( \frac{4\pi}{5}  \right)^2
(2 \ell + 1) \int P(k)
\left[ \frac{j_{\ell}(k)}{k} \right]^2 {\rm d}k,
\end{equation}

with $j_{\ell}$ the spherical Bessel functions.
For
a cutoff radius $x_0$ lying in the range
from 20 to 100 Mpc the location of the peak of $P(k)$ goes from
$k_{\rm peak} = 1.3 \times 10^3$ to $256$.
In our
calculations the shape of $P(k)$
around the maximum
is not so important for the determination of the
anisotropies of the CMB, other than providing an
overall normalization.

\section{Results}

The possibility of using the DMR angular correlation data
to find restrictions on
the spectrum of PDF was exploited here by fitting
the theoretical angular correlation function $C(\theta,3.1^{\circ};x_0,n)$
{\it without any arbitrary normalization factors}
to that obtained by COBE-DMR. For each value of
$\gamma_2$ the parameters
resulting from the fit are the cutoff radius $x_0$
and the spectral index $n$ of the long-wave side of PDF spectrum.

A numerical minimization of the
$\chi^2$ function  using CERN's MINUIT package
(James \& Roos 1975)
gives the results reported in Table 1.

\begin{table}
\caption[ ]{Best fitted parameters}
\begin{flushleft}
\begin{tabular}{llll}
\hline
$\gamma_2$ &   $x_0$~(Mpc)	&  $n$		    &  $\chi^2$   \\
\hline
0.7	   &   $36.4 \pm 12$   &  $0.76 \pm 0.3$  & 64.055 \\
0.8	   &   $34.8 \pm 12$   &  $0.76 \pm 0.3$  & 64.055 \\
0.9	   &   $33.5 \pm 12$   &  $0.76 \pm 0.3$  & 64.055 \\
\hline
\end{tabular}
\end{flushleft}
\end{table}

\begin{figure}
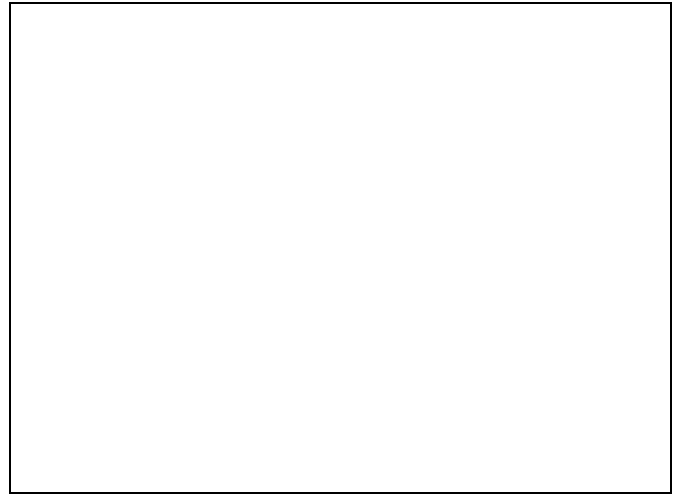

\picplace{6.5cm}
\caption[ ]{Angular power spectrum for the best fit
values as in Fig. 1}
\end{figure}
\begin{figure}
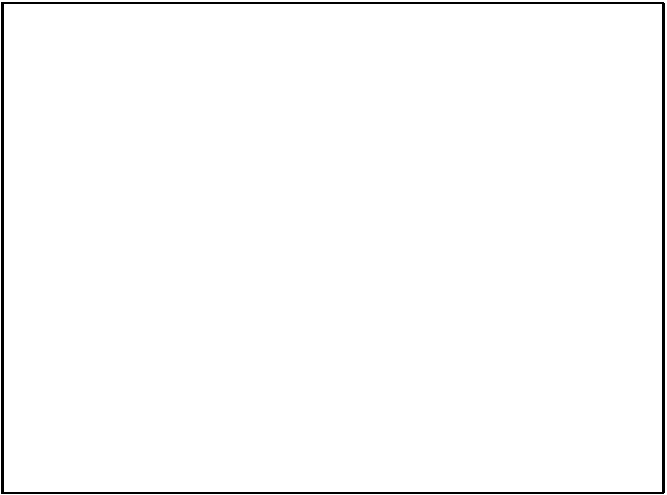

\picplace{6.5cm}
\caption[ ]{Angular correlation function for best fit
values as in Fig. 1 and COBE-DMR data points.
The dipole and quadrupole terms have been removed}
\end{figure}

The fitted parameters are found to be fairly insensitive to
the precise value of
$\gamma_2$, so that we can safely assume that
$x_0 = 35 \pm 12$ Mpc and $n = 0.76 \pm 0.3$.
The errors
quoted  correspond to the 68\% confidence level. The
$\chi^2$ reported are for $69 - 2$ degrees of freedom.
Figures 1, 2 and  3 respectively show the power of density fluctuations, the
CMB angular power spectrum,
and the correlation function
obtained by the fit for the $\gamma_2 = 0.8$ case.
In Fig. 3 the
DMR data is also included in the plot.
Figures 4  and 5 give contour plots of $\chi^2$ near the minimum.
The contours show that the two-parameter fit exhibits
a strong correlation. For $\gamma_2 = 0.8$
we have $x_0 = 0.48 + 45.54n$ Mpc, with $r=0.9996$.
This effect is built-in by the way we normalize
the galaxy-galaxy correlation function. From Eqs. (2) and (3)
it appears that increasing $x_0$ the amplitude of the
galaxy-galaxy correlation function and thereby
of $P(k)$ increases, so that we
need to increase $n$ in order to suppress the excess of long-wave power.
\begin{figure}
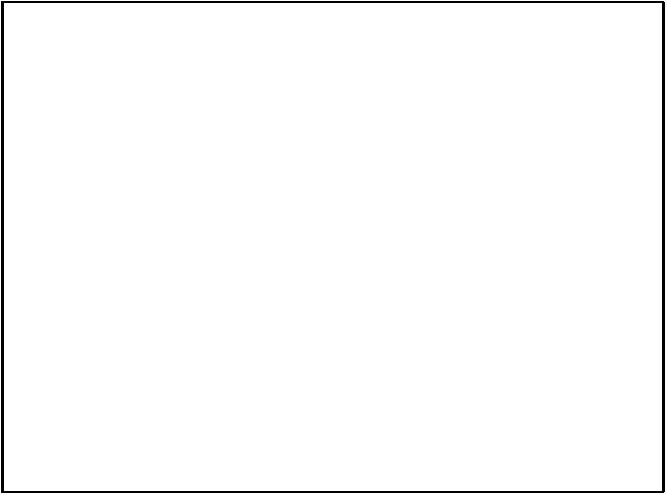

\picplace{6.5cm}
\caption[ ]{Contour plot of $\chi^2$ around the minimum for
$\gamma_2 = 0.8$. The levels are for $\Delta \chi^2 = 2.3,
6.17$ and 11.8, corresponding to 68.3\%, 95.4\% and 99.7\%}
\end{figure}

\begin{figure}
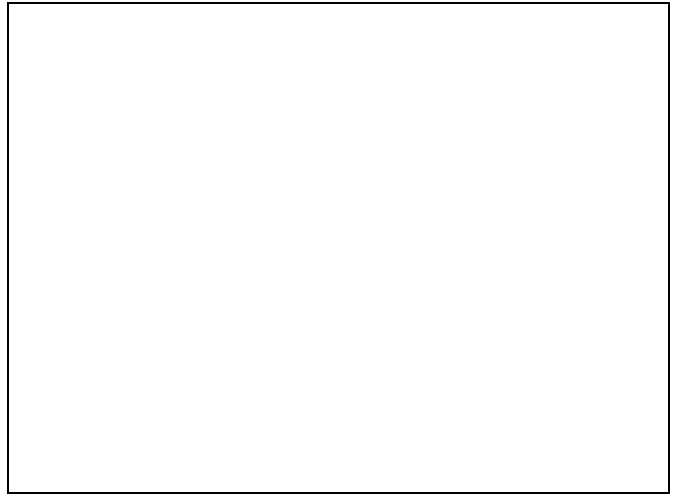

\picplace{6.5cm}
\caption[ ]{Contour plots for a) $\gamma_2=0.7$, and b) 0.9. The
reported
contours are for $\Delta \chi^2 = 2.3$ and 11.8}
\end{figure}

One could argue that if the cutoff radius $x_0$
is an independently well known parameter, then
it should be fixed. Taking $x_0 = 30$ Mpc
(the best value of Calzetti et al. 1992) the
spectral index obtained from the fit is more tightly
constrained: $n=0.64 \pm 0.02$ with a $\chi^2$ of 64.2. This is
far from unity.

Alternatively one could fix $n$ to 1.0 and
ask for the best
$x_0$ that fits the data, still satisfying the second of the conditions above.
This results in $x_0 = 44 \pm 1 h^{-1}$ Mpc
with $\chi^2 = 64.88$. This radius implies
$\lambda_{\rm peak} = 65 h^{-1}$ Mpc.

To compare with the numbers obtained by the COBE-DMR
group we have made the fit using a `pure', unnormalized
power law $P(k) \propto k^n$, and left the quadrupole amplitude
$a_2^2$ as a free parameter of the fit (which in the end
is used to renormalize the spectrum). The results are
$\chi^2 = 64.06$, quadrupole amplitude, $Q_{\rm rms-ps} = (\Delta T)_2
= T_0(a_2^2/4\pi)^{1/2}$, of $16 \pm 8 \; \mu$K and $n = 0.8 \pm 0.3$.
If one fixes $n$ to 1.0 then $\chi^2$ increases to 64.88 while
the quadrupole amplitude decreases to $Q = 15 \pm 5 \; \mu$K.

We have checked the consistency of our results  with
the independent measurement of the RMS quadrupole anisotropy
by COBE-DMR. As can be seen in Fig. 6, the quadrupole evaluated
at the  optimal values of $x_0$ and $n$  agrees with COBE's $Q_{\rm rms-ps}$.

\begin{figure}
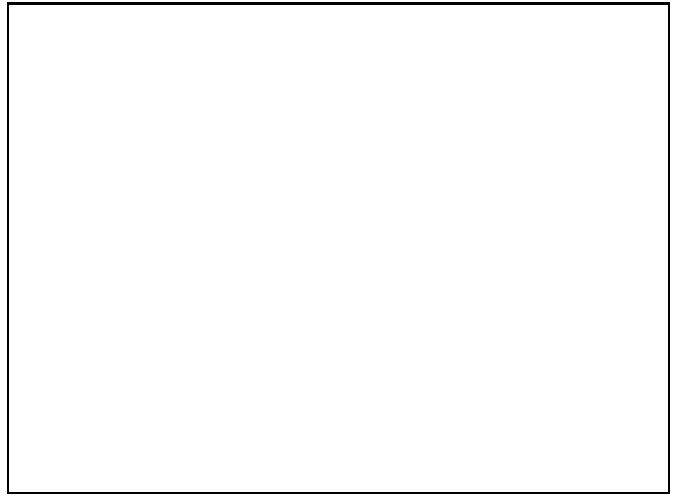

\picplace{6.5cm}
\caption[ ]{Quadrupole amplitudes
derived from the model with $\gamma_2=0.8$ for various values of
$n$.  Solid lines are curves of fixed $n$, with
$n$ = 0.6, 0.7, 0.8, 0.9, 1.0, 1.2 and 1.4 going from the steepest
to the least steep curve.
The quadrupole amplitudes
corresponding to the optimal value of $x_0$ are indicated by squares. The error
bars at the absolute minimum correspond to the 68\% confidence level. The
horizontal lines indicate the 68\% confidence region of
COBE's $Q_{\rm rms-ps}$}
\end{figure}

\section{Discussion}

The cutoff radius $R_0 = 35 \pm 12 h^{-1}$ Mpc agrees very well with
the result found by Calzetti et al. (1992) on the
Zwicky/CfA1,  SSRS, and IRAS PSC catalogs.
Since $R_0$ corresponds to a wavelength of $ 51 h^{-1}$ Mpc it
agrees with the preferred wavelength found by Fabbri and Natale (1993a)
in IRAS PSC, $\lambda = 50 h^{-1}$ Mpc. However,
the best value for the primordial spectral index $n$ is found to
be somewhat lower than
the best value declared by Smoot et al. (1992),
and from that expected from
the scenario of standard inflation.
The possibility of lower values of $n$ from DMR data has been noticed
also by Adams et al. (1992) and Kashlinsky (1992).

Is our analysis consistent with standard inflation? Inspection of our
$\chi^2$ contours in Figs. 4 and 5  shows that it is so only paying
the price of a rather large value of the cutoff radius,
$R_0 \approx 45 h^{-1}$ Mpc. Such a value weakens the agreement with
Calzetti et al. (1992).
However we
should notice that our
model - obtained by the requirement of a minimum perturbation of $\xi(r)$ -
forces the test to be more stringent than other tests on DMR data
since there is no normalization factor as a free
parameter.
We cannot exclude that different spectral shapes might provide results more
favourable towards standard inflation,
also a pure power-law model
with a best-fitted normalization factor
does favor values of $n$ smaller than unity.
It is worth noticing that so small values of $n$ also arise theoretically
from power-law
inflation (Fabbri et al. 1987).

While the best value found for the correlation scale
is completely satisfactory, the peak wavelength found for $P(k)$ is not
consistent with spectra recently derived from redshift surveys on
IRAS galaxies (Fisher et al. 1993), clusters (Einasto et al. 1993) and
both (Jing \&  Valdarnini 1993). There the power spectra appear to
increase towards long wavelenths, and a turnover appears only at the
boundaries of the samples $\lambda \sim 150$  or $180 h^{-1}$ Mpc.
In our view, the significance of such results is weakened by the criticism
of Bahcall et al. (1993), who show that very large distortions must
be generated by  large-scale peculiar velocities. The ratio of the
values of the power spectrum at $\lambda \geq 100 h^{-1} $ and
$\lambda \leq 10 h^{-1}$ Mpc, when calculated in redshift
space, may be $\sim 10$ times  higher than the `true' ratio
in real space ! Unfortunately the recovery of the genuine, real-space
$P(k)$ is model-dependent there. The role of peculiar velocities
is not significant in our work, as well as in Fabbri and Natale
(1993a, b). Further,
it is not clear
whether spectra peaked around $150 h^{-1}$ Mpc can be made  consistent
with COBE's data as well as other anisotropy data
available around $1^o$: Kashlinsky's (1992) results seem to imply that
compatibility cannot be achieved for so broad spectra, while the goal
could be reached for sharp but unphysical peaks.
The present paper shows that the peak wavelength must be much smaller
if we wish (i) to make a really satisfactory fitting of COBE's data
and (ii) to minimally perturb the spectrum directly originated
by $\xi(r)$ in the region where it can be measured.

\acknowledgements{We would like to thank G. Smoot for his comments
and for pointing a serious  mistake in the first manuscript.
This work is partially supported by Colciencias of Colombia
under project \# 1204-05-007-90,  by the Agenzia Spaziale Italiana under
contracts \# 91-RS-43 and \# 92-RS-64, by the Italian Ministry of Foreign
Affairs and the Italian Ministry for University and Scientific and
Technological Research}

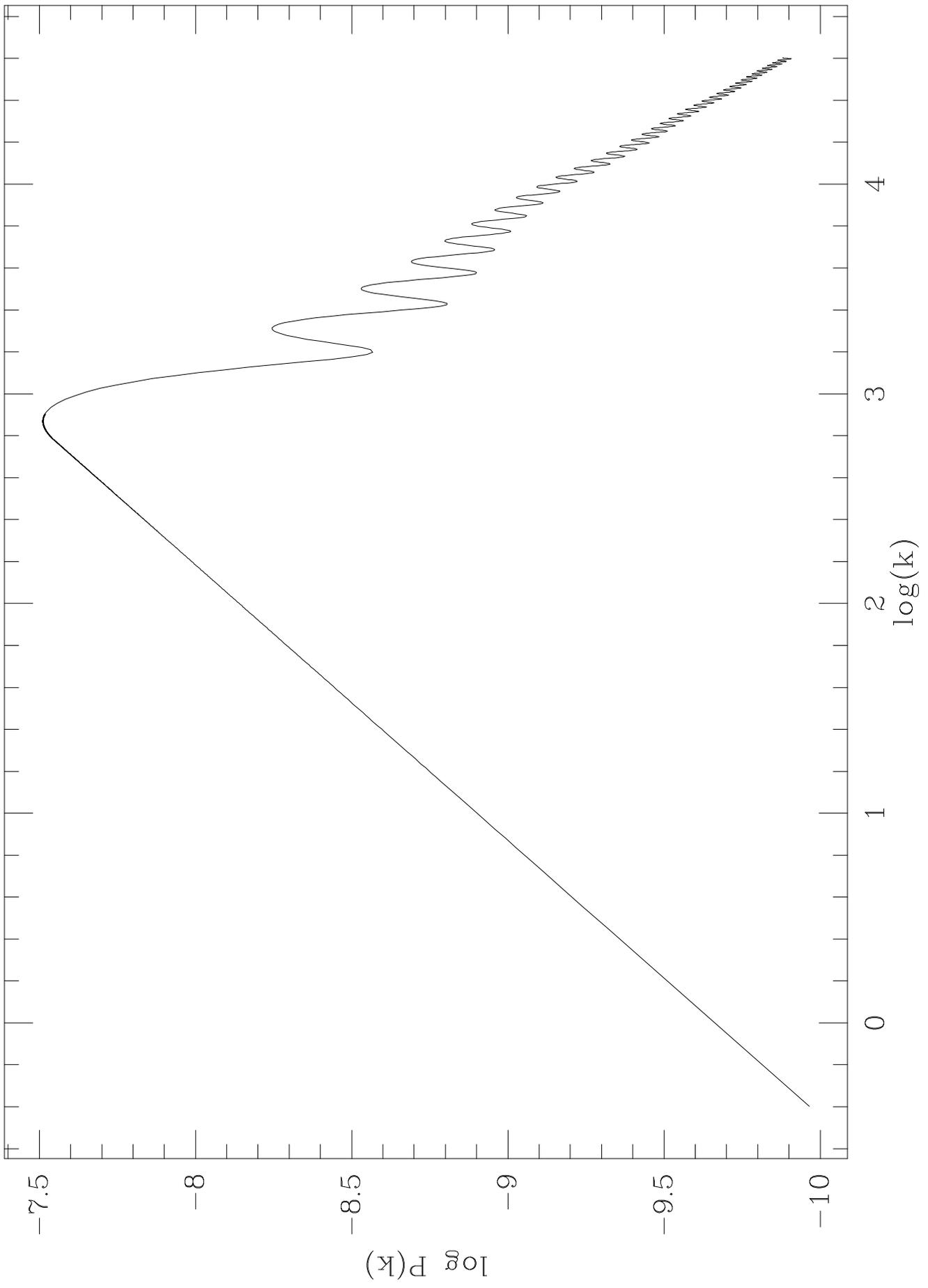
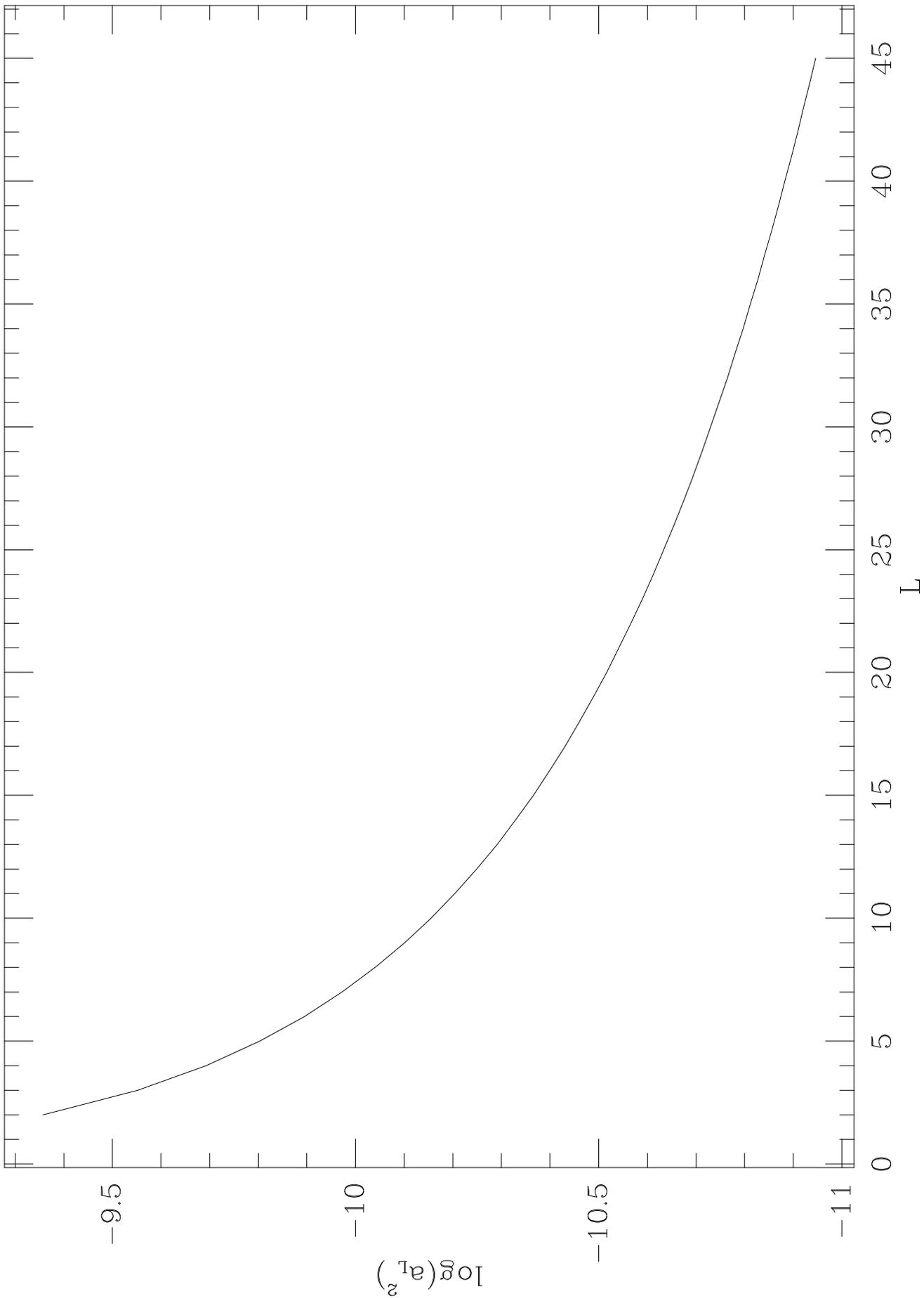
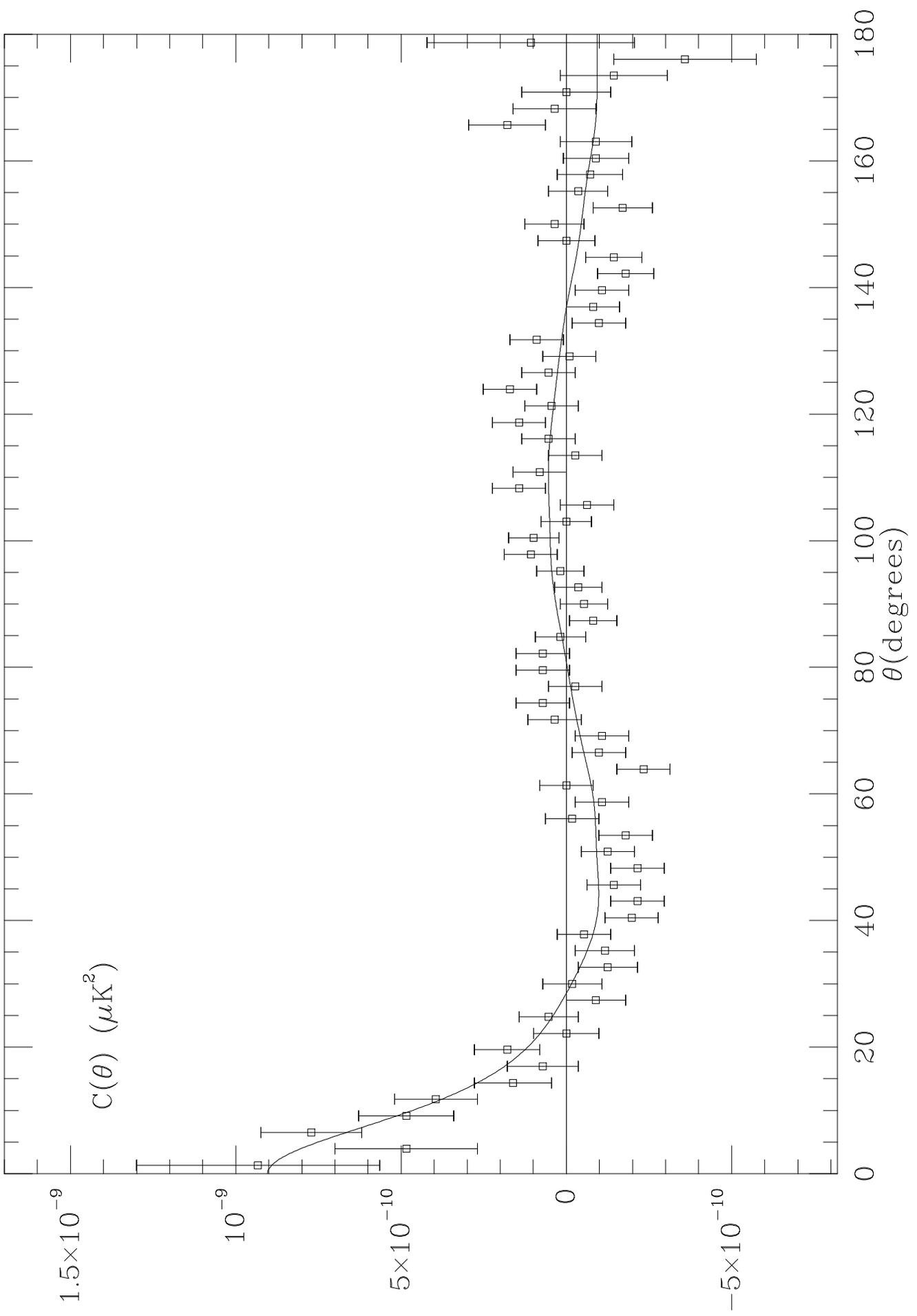
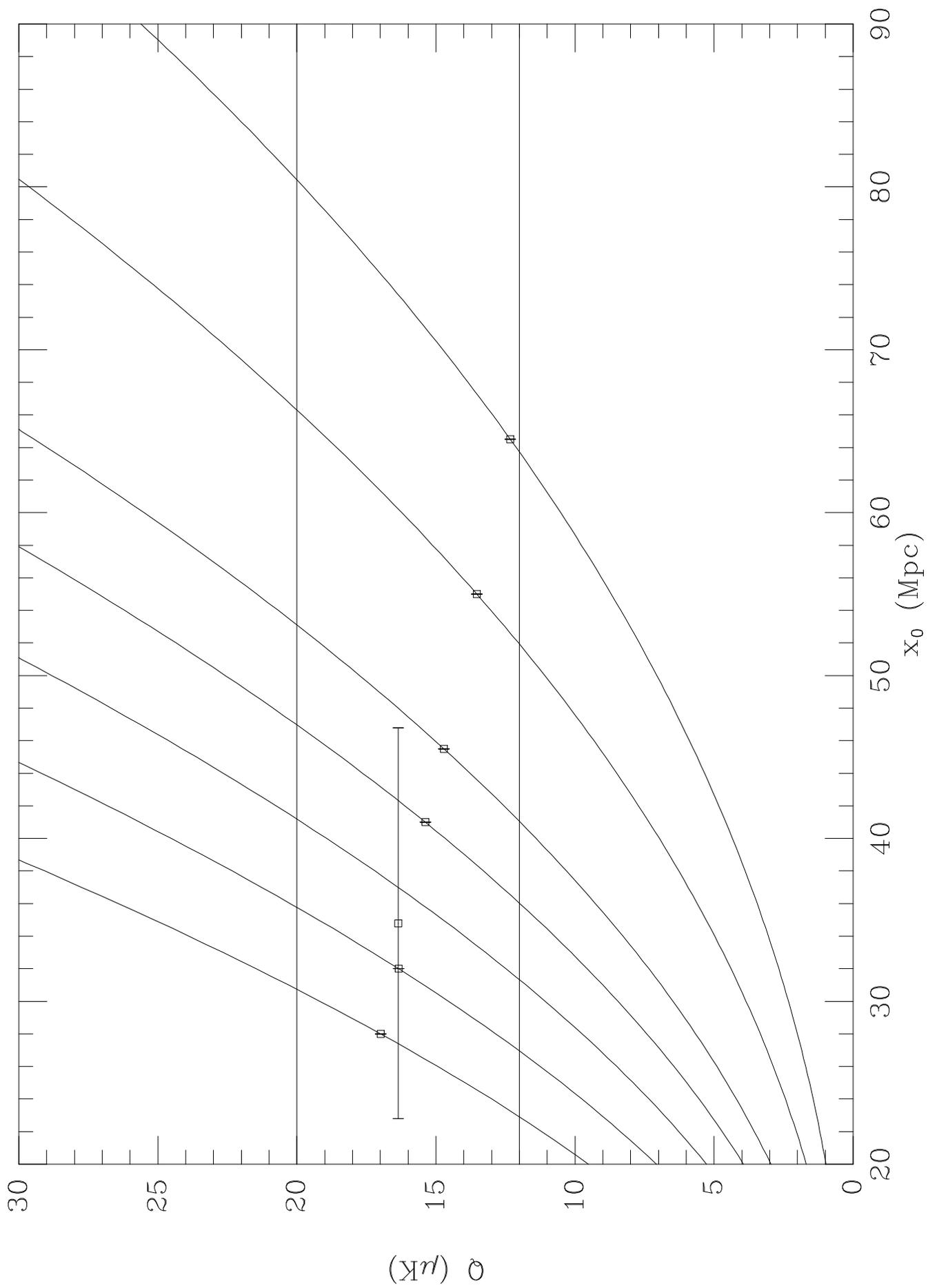
\end{document}